\documentclass[11pt,english,british]{article}
\usepackage{lmodern}
\usepackage[T1]{fontenc}
\usepackage[latin9]{inputenc}
\usepackage{geometry}
\usepackage{color}
\usepackage{amsmath}
\usepackage{amsthm}
\usepackage{amssymb}
\usepackage{stackrel}
\usepackage{graphicx}
\usepackage{setspace}
\usepackage[authoryear]{natbib}
\makeatletter

\numberwithin{equation}{section}
\numberwithin{figure}{section}

\usepackage{lineno}

\makeatother

\usepackage{babel}
\begin{document}

\title{\textbf{Geostatistical inference in the presence of geomasking:
a composite-likelihood approach}}

\author{Claudio Fronterrè\textsuperscript{1}, Emanuele Giorgi\textsuperscript{2},
Peter Diggle\textsuperscript{2}\bigskip\\\textsuperscript{1}Department
of Statistical Sciences, University of Padua, Padua, Italy\\\textsuperscript{2}CHICAS,
Medical School, Lancaster University, Lancaster, United Kingdom }
\maketitle
\begin{abstract}
In almost any geostatistical analysis, one of the underlying, often
implicit, modelling assumptions is that the spatial locations, where
measurements are taken, are recorded without error. In this study
we develop geostatistical inference when this assumption is not valid.
This is often the case when, for example, individual address information
is randomly altered to provide privacy protection or imprecisions
are induced by geocoding processes and measurement devices. Our objective
is to develop a method of inference based on the composite likelihood
that overcomes the inherent computational limits of the full likelihood
method as set out in \citet{Fanshawe2011-be}. Through a simulation
study, we then compare the performance of our proposed approach with
an N-weighted least squares estimation procedure, based on a corrected
version of the empirical variogram. Our results indicate that the
composite-likelihood approach outperforms the latter, leading to smaller
root-mean-square-errors in the parameter estimates. Finally, we illustrate an application of our method to analyse data on malnutrition from a Demographic and Health Survey conducted in Senegal in 2011, where locations were randomly perturbed to protect the privacy of respondents. \textbf{\bigskip{}
\\Keywords: }composite likelihood, geomasking, geostatistics, positional
error.
\end{abstract}

\section{Introduction}

Spatial variation is an important features of many phenomena in science, and
 geo-referenced data are now widely available in the form of measurements
$Y_i$ at locations $x_i$ in a region of interest $A$, commonly called
{\it geostatistical data}. However, an aspect often
neglected is the positional accuracy of the  
measurement locations $x_i$. 
Several empirical studies suggest that positional errors can be neither
random nor negligible
\citep{Dearwent2001-hi,Bonner2003-pf,Cayo2003-th,Rushton2006-ye,Kravets2007-na,Zinszer2010-bv}. Here, we identify
three different kinds of positional error. 

The first kind results from
 the use of imprecise recording devices, as in the case of data collected using GPS receivers
or satellites. Several factors, including the height at which the device is placed,
air transparency and clouding, can  affect the precision of recorded coordinates \citep{Devillers2006-in}. 

The second kind arises when measurement locations
cannot be released because of the need to preserve confidentiality. Random or deterministic perturbation of the locations is then applied
in a  process known as {\it geomasking} \citep{Armstrong1999-lz}. \par

The third kind, known as {\it geocoding}, corresponds to the process
 of converting text-based
addresses into geographic coordinates. Imprecision is
 then introduced due to incorrect 
 placement along a street segment \citep{Zandbergen2009-yr} or,  more generally, 
because postcode systems typically assign the same geocoded location to multiple
addresses.
 \par

\citet{Jacquez2012-qq} outlines a research agenda whose objective is to develop a rigorous methodological framework that deals with positional error. Ignoring positional error can lead to invalid inferences, including biased estimates of 
diseases rates 
\citep{Zimmerman2006-zm,Zimmerman2007-fo,Goldberg2012-rl}, exposure
effects \citep{Zandbergen2007-gf,Mazumdar2008-pk} and spatial
covariance  parameters \citep{Gabrosek2002-ws,Arbia2015-qm}.
 It can also impair the performance of
cluster detection algorithms \citep{Jacquez2000-uc,Zimmerman2010-cu} and
tests for space-time interaction \citep{Malizia2013-jc}.  \citet{Jacquez2012-qq} states that most public health studies ignore 
the issue of geocoding spatial uncertainty because of a lack of principled statistical methods for dealing with it. \par

\citet{Gabrosek2002-ws} examine how positional uncertainty in the measurement locations $x_i$ of a geostatistical data-set can affect
estimation of the covariance structure
 of a stationary and isotropic Gaussian process $S(x)$ and show how to account for 
this by correcting the conventional
kriging equations used for spatial prediction. They find that their corrected  approach performs better than ordinary
kriging and that the presence of positional error inflates both the bias and the mean squared prediction error of ordinary kriging. \citet{Cressie2003-es} propose a more general approach to deal with the effects of positional error on the mean component of geostatistical models and apply this
 to analyse remote sensing data on total column ozone,
where positional error is caused by allocation of each measured
value to the centre of the nearest grid-cell. \citet{Fanshawe2011-be}
propose a model-based solution by
deriving the likelihood function for a linear Gaussian geostatistical
model  incorporating positional error. They also allow for
positional error in a notional prediction location $x$ and find that  this induces
skewness into the predictive distribution of $S(x)$.
However, the authors report that 
the computational burden of their method 
 makes it infeasible even for moderately large geostatistical data-sets. \par

In this paper we focus our attention on positional error due to geomasking.
 In Section 2, we provide more details on geomasking, 
derive the parametric form of the theoretical variogram in the presence of
geomasking and propose a method of variogram-based parameter estimation. 
In Section 3, we derive the likelihood function as in \citet{Fanshawe2011-be} and propose an approximation using composite likelihood
 \citep{Varin2011-da}.
 In Section 4, we conduct a simulation study that compares the performance of the
variogram-based and composite likelihood estimators.
In Section 5, we analyse data on malnutrition from 
 from a Demographic Health Survey conducted in Senegal in 2011.
Section 6 is a concluding discussion. 

\section{Geomasking and its effects on the spatial covariance structure}

Geomasking
consists of adding a stochastic or deterministic displacement to the spatial
coordinates $x_i$ of a geostatistical data-set.
\citet{Armstrong1999-lz} advocated geomasking as an improvement on the standard practice
of aggregating health records to preserve the confidentiality of 
information about individuals  who might otherwise be identified
by their exact spatial location.

Here, we consider stochastic perturbation methods, as these are the most commonly used in practice. For example, the Forest Inventory Analysis Program \citep{McRoberts2005-qu}
the Living Standard Indicator Survey \citep{Grosh1996-wh} and the
Demographic and Health Surveys \citep{Burgert2013-oq} have adopted this approach to
allow public sharing of data while protecting respondents' confidentiality. Other geomasking techniques have been proposed, such as \textit{doughnut} geomasking \citep{Hampton2010-nd}
and Gaussian bimodal displacement \citep{Cassa2006-nl}. However, these are less
used in practice because they introduce excessive bias without significant 
reduction in the risk of identification. For a thorough review on geomasking
methods, see \citet{Zandbergen2014-jt}. \par

Let $Y_{i}$ denote the random variable associated with the outcome of interest, measured at locations $x_{i}$, for $i=1,\ldots,n$. Let $S(x)$  be a stationary, isotropic  Gaussian process with mean zero, variance $\sigma^2$
and \citet{Matern1960-ou} correlation function, given by
\[
\rho(u_{ij};\phi,\kappa)=\{ 2^{\kappa-1}\Gamma(\kappa)\} ^{-1}(u_{ij}/\phi)^{\kappa}K_{\kappa}(u_{ij}/\phi),
\]
where $u_{ij} = \|x_{i}-x_{j}\|$ is the Euclidean distance between any two locations $x_{i}$ and $x_{j}$, $\phi>0$ is a scale parameter,
$\kappa>0$ is a shape parameter that regulates the smoothness of $S(x)$ and $K_{\kappa}(\cdot)$
denotes the modified Bessel function of order $\kappa$. 
Also, let be $Z_{i}$ a set of mutually
independent  ${\rm N}(0,\tau^2)$ random variables. 
The standard linear geostatistical model \citep{Diggle2007-ri} then  takes the form 
\begin{equation}
Y_{i}=d(x_{i})^{\top} \beta+S(x_{i})+Z_{i}:i=1,\ldots,n.
\label{eq:4.0}
\end{equation}
where, for any location $x$,
 $d(x)$ is a vector of explanatory variables with regression coefficients $\beta$.

The {\it  variogram} for the outcome $Y_{i}$ is defined as 
\begin{eqnarray}
V(u_{ij}) & = & \frac{1}{2}{\rm Var}\{Y_{i} - Y_{j}\} \nonumber \\
& = & 
\frac{1}{2} {\rm E}[\{(Y_{i}-d(x_{i})^\top\beta)-(Y_{j}-d(x_{j})^\top\beta)\}^2]
\label{eq:4.00}
\end{eqnarray}
When $S(x)$ is stationary and isotropic, \ref{eq:4.00} reduces to 
$$
V(u_{ij})=\tau^{2}+\sigma^{2}\{ 1-\rho(u_{ij})\}. 
$$
\par

In the presence of positional error, the true location is an
unobserved random variable, which we denote by $X_{i}^*$. 
We observe the realised  value of the displaced location,
\begin{equation}
X_{i}=X_{i}^{*}+W_{i},\label{eq:4.1}
\end{equation}
where the $W_{i}$ represent the positional
error process. We assume that the $W_i$  are  mutually independent random variables
whose bivariate density is symmetric about the origin with variance matrix $\delta^2 I$;
we call   $\delta^{2}$ the {\it positional error variance}. In what follows
we will assume  
that $W_{i}$ follows a Gaussian distribution, but the results in the remainder
of this Section 
 hold for any other symmetric distribution.

Let $U_{ij}=\|X_{i}-X_{j}\|$ and
 $V_{ij}=
\{(Y_{i}-d(x_{i})^{\top} \beta) - (Y_{j}-  d(x_{j})^{\top} \beta)\}^{2}/2$. 
It follows from (\ref{eq:4.0}) and (\ref{eq:4.1}) that $U_{ij}$ and $V_{ij}$
are conditionally independent given $U_{ij}^{*}=\|X_{i}^*-X_{j}^*\|$. 
Using the notation
$[\thinspace \cdot \thinspace]$ to mean
 ``the distribution of''  it then follows that
\begin{equation}
[V_{ij}\mid U_{ij}]=\int_{0}^{\infty}[V_{ij} \mid  U_{ij}^{*}][U_{ij}^{*}\mid u_{ij}]d U_{ij}^{*}.\label{eq:4.2}
\end{equation}
Also,  $[V_{ij}\mid U_{ij}^{*}] = V_{Y}(U_{ij}^{*})\chi_{(1)}^{2}$
and $[U_{ij}^{*}\mid u_{ij}]$ follows a \citet{Rice-_Bell_Labs_Technical_Journal1944-fp} distribution with scale parameter $\sqrt{2}\delta$ which we denote as $Rice(u_{ij},\sqrt{2}\delta)$; we give more details on the Rice distribution in the appendix.
Taking the expectation of (\ref{eq:4.2}) with respect to 
$[Y_{i}, Y_{j} | u_{ij}]$ gives the theoretical variogram 
\begin{equation}
V_Y(u_{ij})=\tau^{2}+\sigma^{2}\{ 1-E[\rho (U_{ij}^{*}) \mid u_{ij} ]\},
\label{eq:4.3}
\end{equation}
where $E[\cdot]$ denotes expectation with respect to $U_{ij}^*$.
As $\delta \rightarrow 0$, 
$V_Y(u_{ij})$ converges
to the true variogram $V(U_{ij}^{*})$ given by (\ref{eq:4.00}), whereas
as  $\delta \rightarrow\infty$
the spatial correlation structure of the data is destroyed and
$V_Y(u_{ij}) \rightarrow \tau^2 + \sigma^2$. \par

\begin{figure}[h]

\centering
\includegraphics[scale=0.7]{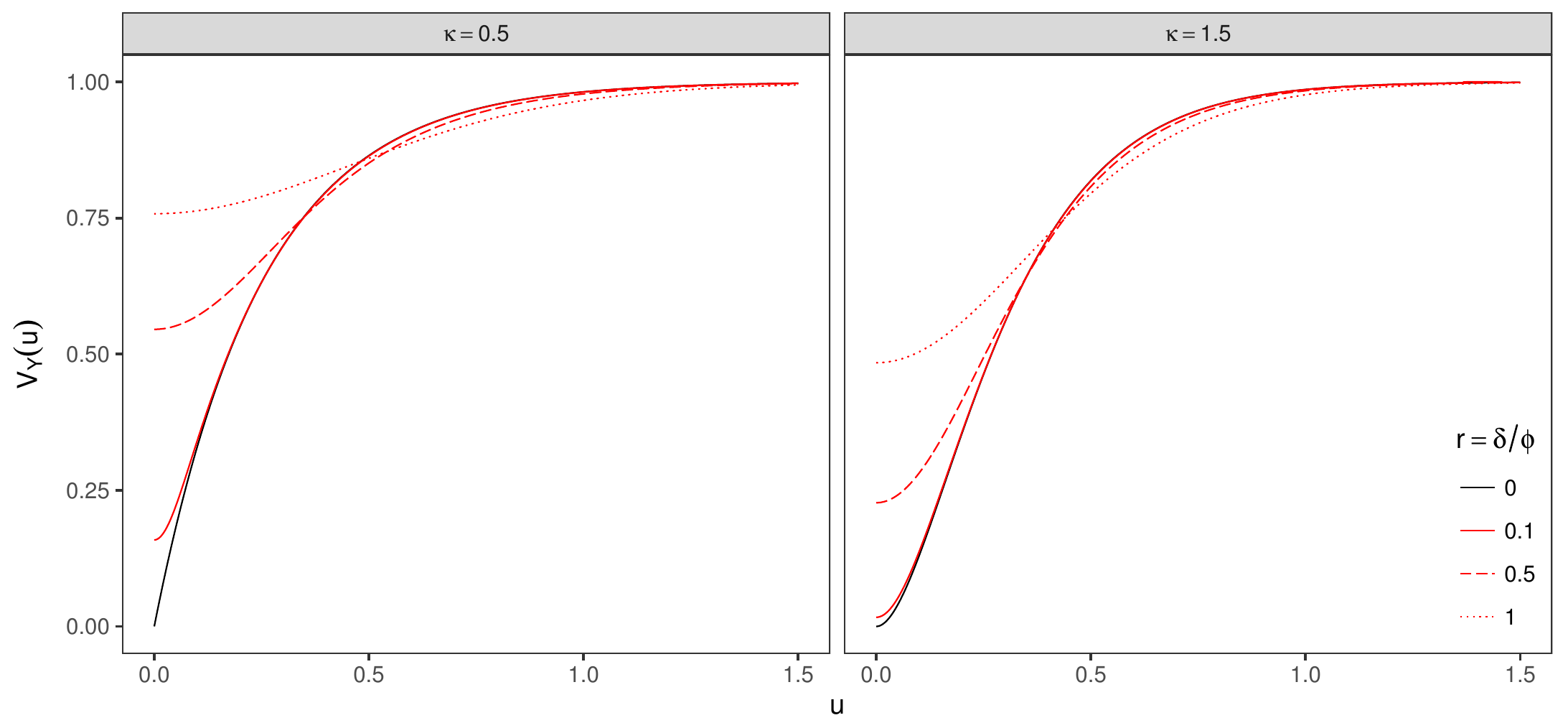}
\caption{\label{fig:Departures-(red-lines)}Departures (red lines) from the
true variogram (solid black line) with $\sigma^{2}=1$ and $\tau^{2}=0$
for increasing values of $r=\delta/\phi$. Mat\'ern correlation functions
with two different shape parameters are used. }
\end{figure}

In \eqref{eq:4.3}, the expectation on the right-hand side is not 
generally available in closed form. An exceptional case is
the Gaussian correlation function, 
$\rho(u_{ij})=\exp\{ -(u_{ij}/\phi)^2\}$, 
which is the limiting case of the Mat\'ern 
correlation function as $\kappa \rightarrow \infty$. In this case,
\begin{equation}
E[\rho(U_{ij}^{*}) \mid u_{ij}]=\frac{1}{1+(2r)^{2}}\exp\left\{ -\left(\frac{u_{ij}}{\phi\sqrt{1+(2r)^{2}}}\right)^{2}\right\} ,\label{eq:4.4}
\end{equation}
where $r=\delta/\phi$. Hence, the magnitude
of the bias in variogram estimation
 induced by geomasking depends on the ratio between the
standard deviation of the positional error distribution and the range parameter of 
the correlation function of $S(x)$.
  Additionally, as $u_{ij} \rightarrow 0$ in (\ref{eq:4.4}),
$E[\rho(U_{ij}^{*}) \mid u_{ij} ] \rightarrow \{ 1+(2r)^{2}\} ^{-1} < 1$ and, as $r\rightarrow \infty$, $E[\rho(U_{ij}^{*}) \mid u_{ij} ] \rightarrow 0$.
We conclude that for variogram estimation, the main effect of
ignoring geomasking is to introduce bias into the estimates of $\tau^2$ and $\phi$. Figure \ref{fig:Departures-(red-lines)} shows departures from the true variogram (black line) for different levels of $r$ when the true correlation function is Mat\'ern. We observe that as $r$ increases, both the scale of the spatial correlation and  the discontinuity at the origin increase, yielding a variogram that is flatter
overall.  \par

We now use  (\ref{eq:4.3}) to develop a positional-error-corrected 
method of variogram-based  n-weighted least squares
covariance parameter estimation. We assume the positional error variance $\delta^2$ to be known, as
this  is a necessary requirement for identifiability of the model parameters.

 Let $\theta$ denote
the vector of covariance parameters to be estimated; typically, $\theta=(\tau^2,\sigma^2,\phi)$.
Write $r_i = y_i - d(x_i)^\top \tilde{\beta}$, where $\tilde{\beta}$ is a
preliminary estimate of $\beta$, for example the ordinary least squares regression
estimate,  $v_{ij} = \frac{1}{2}(r_i - r_j)^2$ and $u_{ij} = \|x_i - x_j\|$.
Our objective function is
\begin{equation}
{\cal F}_{n}(\theta)=\stackrel[{\scriptstyle k=1}]{{\scriptstyle m}}{\sum}n_{k}\{ v_{k}-V_{Y}(u_{k};\theta)\} ^{2},
\end{equation}
where: $v_{k}$ are the sample variogram ordinates, obtained by averaging
 all $v_{ij}$ such $(k-1)h<u_{ij}\leq k h$, where $h$
is the bin width; $u_{k}=(k-0.5)h$ is the mid-point
of the corresponding bin interval; $n_{k}$
is the number of pairs $(i,j)$ points 
that contribute to $v_{k}$; and $V_{Y}(u_{k};\theta)$ is the corrected variogram as defined by \eqref{eq:4.3}.

\section{Likelihood-based inference for the linear Gaussian model}

To derive the likelihood function for 
the linear geostatistical model with positional error, we use the following notation:
 $Y = (Y_{1}, \ldots, Y_{n})$ is the collection of all the random variables associated with our outcome of interest; $S = (S(x_{1}), \ldots, S(x_{n}))$ is the vector of the spatial random effects at the observed locations $x_{i}$ for $i=1,\ldots,n$; $X = \{X_{1},\ldots,X_{n}\}$ and $X^* =\{X_{1}^*,\ldots, X_{n}^*\}$ are the perturbed and the true locations, respectively. We then factorize their joint distribution as 
\begin{eqnarray*}
	\left[Y,S,X,X^{*}\right] & = & \left[Y\mid S,X,X^{*}\right]\left[S,X,X^{*}\right]\\
	& = & \left[Y\mid S,X^{*}\right]\left[S\mid X,X^{*}\right]\left[X,X^{*}\right]\\
	& = & \left[Y\mid S,X^{*}\right]\left[S\mid X^{*}\right]\left[X^{*}\mid X\right]\left[X\right],
\end{eqnarray*}
where: $\left[Y\mid S,X^{*}\right]= \prod_{i=1}^n \left[Y_i\mid S(X_i^{*})\right]$;
 $\left[Y_i\mid S(X_i^{*})\right]$ is Gaussian distribution with
 mean $S\left(X_{i}^{*}\right)$ and variance $\tau^{2}$; $\left[S\mid X^{*}\right]$ is multivariate Gaussian with mean
zero and covariance matrix $\Sigma$ such that $[\Sigma]_{ij} = \sigma^2 \rho(U^*_{ij}; \phi, \kappa)$; and $\left[X_{i}^{*}\mid X_{i}\right]$ is a bivariate Gaussian distribution with mean $X_{i}$ and covariance matrix $\delta^{2}I_{2}$. Also, note that in
the above equation $\left[Y\mid S,X,X^{*}\right]=\left[Y\mid S,X^{*}\right]$
because of the conditional independence between $Y$ and $X$
given $X^{*}$. \par 

The likelihood function for the unknown vector of parameters $\psi=\left(\beta, \sigma^{2},\phi,\tau^{2}\right)$  is 
\begin{eqnarray}
L\left(\psi\right) & = & \left[Y,X; \psi\right]\nonumber \\
& = & \int\int\left[Y,X,X^{*},S;\psi\right]dSdX^{*}\nonumber \\
& = & \int\int\left[Y\mid X,X^{*},S;\psi\right]\left[S,X,X^{*};\psi\right]dSdX^{*}\nonumber \\
& = & \int\int\left[Y\mid X^{*},S;\psi\right]\left[S\mid X^{*};\psi\right]\left[X^{*}\mid X,\right]\left[X\right]dSdX^{*}\nonumber \\
& \propto & \int\int\left[Y\mid X^{*},S;\psi\right]\left[S\mid X^{*};\psi\right]\left[X^{*}\mid X\right]dSdX^{*},\label{eq:inference}
\end{eqnarray}
After integrating out $S$ in  (\ref{eq:inference}), the final expression 
for the likelihood is
\begin{equation}
L(\psi) \propto \int\left[Y\mid X^{*}\right]\left[X^{*}\mid X\right]dX^{*},\label{eq:likelihood}
\end{equation}
where $\left[Y\mid X^{*},\psi\right]$ is a multivariate Gaussian distribution with mean $D^*\beta$ and covariance matrix $\Sigma + \tau^2 I_{n}$; here, $D^*$ denotes the matrix of covariates at the true locations $X^*$. \par

\citet{Fanshawe2011-be} propose to approximate (\ref{eq:likelihood})
by Monte Carlo integration. Given $\psi$ and $\delta$,
they draw $B$ independent
samples from $\left[X^{*}\mid X;\psi\right]$. 
The resulting approximation to the likelihood is then obtained as
$$
L(\psi)\approx \frac{1}{B} \left[Y\mid X_{(b)}^{*};\psi\right]
$$
where $X_{(b)}^{*}$ is the $b$-th samples from $\left[X^{*}\mid X;\psi\right]$. Maximization of the above expression is computationally intensive since a single evaluation of the approximated likelihood has a computational burden of order $O\left(B\times n^{3}\right)$. For this reason, Fanshawe and Diggle conclude that reliable computation of the standard errors for the maximum likelihood estimates is infeasible.

\subsection{Composite likelihood}
\label{subsec:cl}
We propose to approximate the likelihood in \eqref{eq:likelihood}
using the composite likelihood method. The resulting estimating equation obtained from
the derivative of the composite log-likelihood is an unbiased estimating
equation \citep{Varin2011-da}. This approach has been applied to
standard geostatistical models to make computations faster when the
number of spatial locations is demanding (\citealt{Vecchia1988-zq,Hjort1994-aw,Curriero1999-ah,Stein2004-lh,Caragea2006-xp,Caragea2007-vg,Mateu2007-sb,Bevilacqua2012-ic,Bevilacqua2015-rg}). \par
The resulting approximation is obtained by treating as independent each of the pairs of bivariate densities, to give
\begin{eqnarray}
\log L(\psi) \approx \log L_{CL}(\psi) &=&\stackrel[{\scriptscriptstyle i=1}]{{\scriptscriptstyle n-1}}{\sum}\stackrel[{\scriptscriptstyle j=i+1}]{{\scriptscriptstyle n}}{\sum}\log [Y_{i},Y_{j};\psi] \nonumber \\
&=& \stackrel[{\scriptscriptstyle i=1}]{{\scriptscriptstyle n-1}}{\sum}\stackrel[{\scriptscriptstyle j=i+1}]{{\scriptscriptstyle n}}{\sum}\log\int_{0}^{\infty}\left[Y_{i},Y_{j}\mid U_{ij}^{*}\right]\left[U_{ij}^{*}\mid u_{ij}\right]dU_{ij}^{*}.
\label{eq:l1}
\end{eqnarray}

Hence, computation of the approximate likelihood requires the integration of $n\left(n-1\right)/2$ univariate integrals. Additionally, note that as the distance between a pair of observations increases,
the density function of $[Y_{i}, Y_{j} | U_{ij}^*]$ tends to
\[
[Y_{i}][Y_{j}] = \frac{1}{2\pi\left(\sigma^{2}+\tau^{2}\right)}\exp\left\{ -\frac{y_{i}^{2}+y_{j}^{2}}{2\left(\sigma^{2}+\tau^{2}\right)}\right\} .
\]
We exploit this fact to further approximate the likelihood function as
\begin{equation}
[Y_{i}, Y_{j} | U_{ij}^*] \approx [Y_{i}, Y_{j} | U_{ij}^*] I\left(u_{ij} \leq t\right)+[Y_{i}][Y_{j}]I\left(u_{ij}>t\right),
\label{eq:approx}	
\end{equation}
where $I\left(\mathcal{P}\right)$
is an indicator function which takes the value 1 if property $\mathcal{P}$ is verified and 0 otherwise, whilst $t$ is a threshold such that all pairs of observations that are more than
a distance $t$  apart are assumed to be independent. Let $U^*_{ij,(b)} = Q(s_{b}; u_{ij}, \delta)$ for $b=1,\ldots,B$, where $Q(\cdot; u_{ij}, \delta)$ is the quantile function of a $Rice(u_{ij}, \delta)$ and $s_b$ is the $b$-th term of a \citet{Halton1960-bc} sequence. \par

 For $u_{ij} \leq t$, we compute the univariate integral in \eqref{eq:l1} using quasi Monte Carlo methods, i.e.
$$
\int_{0}^{\infty}\left[Y_{i},Y_{j}\mid U_{ij}^{*}\right]\left[U_{ij}^{*}\mid u_{ij}\right]dU_{ij}^{*} \approx \frac{1}{B} \sum_{b=1}^B [Y_{i}, Y_{j} | U^*_{ij,(b)}].
$$
If $u_{ij} > t$, the integral equals $[Y_{i}][Y_{j}].$

\subsection{Uniform geomasking}
\label{subsec:uniform_geomasking}

A commonly used
alternative to Gaussian geomasking is uniform geomasking. 

Let $W = (W_{1}, W_{2})$; we now define the positional error process as
\begin{equation}
\begin{cases}
W_{1} = R \cos \Lambda \\
W_{2} = R \sin \Lambda
\end{cases},
\label{eq:unif-geomask}
\end{equation}
where $R$ and $\Lambda$ are two independent uniform random variables in $[0,\delta]$, with $\delta$ now denoting the maximum displacement distance, and $[0, 2\pi]$, respectively. However, note that the resulting distribution of $W$ is not uniform within a disc of radius $\delta$ but has a higher probability density toward the center of the disc. Under uniform geomasking $[U_{ij}^* | u_{ij}]$ is an intractable distribution, making computation of the likelihood function in \eqref{eq:l1} cumbersome. \par

In the application of Section \ref{sec:application}, we propose to approximate $[U_{ij}^* | u_{ij}]$ under uniform geomasking with a $Rice(u_{ij}, \delta/\sqrt{6})$ since the variance for each of the components of $W$ in \eqref{eq:unif-geomask} is $\delta^2/6$. \par

We illustrate the quality of this approximation as follows. 
We first express $U_{ij}^*$ in terms of $R$, $\Lambda$ and $u_{ij}$ as
\begin{equation}
\label{eq:u_ij_star}
U_{ij}^* = \sqrt{u_{ij}^2+R^2-2u_{ij}R\sin \Lambda}.
\end{equation}
We then simulate 100,000 samples from the uniform distribution on $[0,\delta]$
and 100,000 samples from the uniform distribution on $[0,2\pi]$. For a given value of 
$u_{ij}$, we then compute the empirical cumulative density function (CDF) 
of the corresponding 100,000 values
generated from $[U_{ij}^* | u_{ij}]$, based on \eqref{eq:u_ij_star}. \par

\begin{figure}
\begin{center}
\includegraphics[scale=0.8]{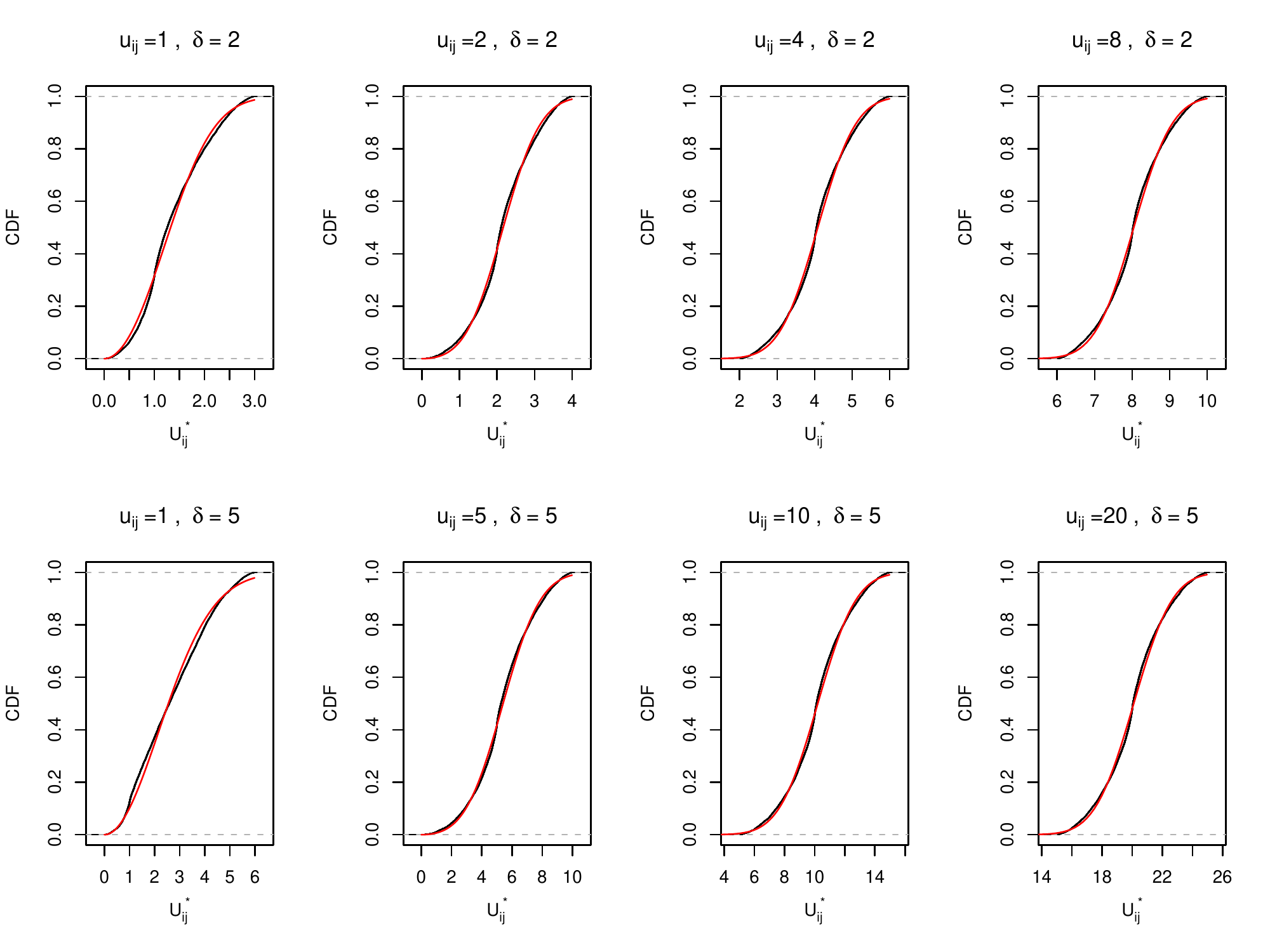}
\caption{Each plot shows the empirical cumulative density function (CDF) based on $100,000$ samples generated from $[U_{ij}^* | u_{ij}]$ under uniform geomasking (black line) and the CDF of a $Rice(u_{ij}, \delta/\sqrt{6})$ (red line). The corresponding values of $u_{ij}$ and $\delta$ are shown in the heading of each plot. \label{fig:approx_ug}}
\end{center}
\end{figure}

Figure \ref{fig:approx_ug} reports the result of the simulation.
The discrepancies between the empirical CDF under uniform geomasking (black line) and the CDF of a $Rice(u_{ij}, \delta/\sqrt{6})$ are small in all of the eight scenarios considered.
We used values of $\delta=2$ and $\delta=5$ as these  correspond to the forms
of geomasking 
that were used in the application of Section \ref{sec:application}.

\section{Simulation study}

We conducted a simulation study to quantify the effects of positional
errors on parameter estimation as follows.
\begin{enumerate}
	\item Generate $n=1000$ locations  $[X^*]$  as a homogeneous Poisson process over the square $[0, 15]\times[0,15]$.
	\item Simulate the outcome data from $[Y | X^*]$ as indicated in \eqref{eq:4.0}, setting $\beta=0$.
	\item Simulate from $[X | X^*]$ using Gaussian geomasking to obtain $X$.
	\item Estimate $\psi$ to obtain $\hat{\psi}_{i}$ for the $i$-th simulated data-set using:
	\begin{itemize}
		\item variogNaive, a parametric fit to the variogram that ignores using weighted least squares (WLS);
		\item variogAdj, a parametric fit to the variogram that corrects for positional error using WLS;
		\item geoNaive, maximum likelihood estimation under
		a linear geostatistical model that ignores positional error;
		\item CL, the composite likelihood method of Section \ref{subsec:cl};
		\item ACL1, as CL but assuming pairs of observations $Y_{i}$ and $Y_{j}$ to be independent for values of the correlation between $Y_{i}$ and $Y_{j}$ below $0.05$;
		
		\item ACL2, as CL but assuming pairs of observations $Y_{i}$ and $Y_{j}$ to be independent for values of the correlation between $Y_{i}$ and $Y_{j}$ below $0.000005$;
	\end{itemize}
	\item Repeat STEPS 1 to 4  $s=500$ times.
	\item Calculate the bias, $$\frac{1}{s}\sum_{i=1}^{s}\hat{\psi}_{i}-\psi,$$
	and the root-mean-square-error (RMSE),
	$$\sqrt{\frac{1}{s}\sum_{i=1}^{s}\left(\hat{\psi}_{i}-\psi\right)^{2}}.$$
\end{enumerate}
We define the following scenarios: (a) $\sigma^2=1$, $\tau^2=0$, $\kappa=0.5$ and $\phi=0.25$; (b) $\sigma^2=1$, $\tau^2=0$, $\kappa=1.5$ and $\phi=0.16$. In both scenarios, we let $r=\delta/\phi$ vary over the set $\{0.2,0.4,0.6,0.8,1\}$. We report the results in Table \ref{Tab:Gauss1} and \ref{Tab:Gauss2}. As expected, BOTH variogNaive and geoNaive overestimate $\tau^2$ and $\phi$ but underestimate $\sigma^{2}$. However, the estimated total variance is not affected by geomasking. In both scenarios, CL shows the smallest RMSE than the other methods for all parameters. The ACL1 and ACL2 show a slight increase in bias and RMSE but with a considerable gain in computational speed. We also observe that the effects of ignoring positional error are less strong on parameter estimation for $\kappa=1.5$ than for $\kappa=0.5$, as the differences in bias and RMSE between the naive and corrected methods are smaller. 

\begin{table}[h]
	\renewcommand{\arraystretch}{1.15}
	\centering
\caption{\label{Tab:Gauss1}Bias and root-mean-square-error (RMSE) of the parameter estimates from the naive methods (variogNaive and geoNaive) and those accounting for positional error (variogAdj,  CL, ACL1 and ACL2). The true correlation function is Mat\'ern with $\kappa=0.5$.}
	\begin{tabular}{lccccccc}
		\hline
		& \multicolumn{2}{c}{$\sigma^2$} &  \multicolumn{2}{c}{$\phi$} &  \multicolumn{2}{c}{$\tau^2$} & $r=\delta/\phi$ \\ 
		\textbf{Method}  & Bias & RMSE & Bias & RMSE & Bias & RMSE \\ 
\hline
		variogNaive & -0.106 & 0.111 & 0.026 & 0.036 & 0.101 & 0.104 & 0.2 \\ 
		variogAdj & -0.050 & 0.069 & 0.013 & 0.029 & 0.014 & 0.017 & 0.2 \\ 
		geoNaive & -0.166 & 0.196 & 0.037 & 0.039 & 0.149 & 0.179 & 0.2 \\ 
		CL & -0.053 & 0.069 & -0.001 & 0.017 & 0.015 & 0.016 & 0.2 \\ 
		ACL2 & -0.053 & 0.069 & -0.001 & 0.017& 0.015 & 0.016 & 0.2 \\ 
		ACL1 & -0.052 & 0.068 & 0.005 & 0.021 & 0.015 & 0.016 & 0.2 \\ 
		\hline
		variogNaive & -0.266 & 0.327 & 0.070 & 0.120 & 0.279 & 0.319 & 0.4 \\ 
		variogAdj & -0.053 & 0.072 & 0.013 & 0.035 & 0.002 & 0.002 & 0.4 \\ 
		geoNaive & -0.323 & 0.421 & 0.083 & 0.124 & 0.321 & 0.410 & 0.4 \\ 
		CL & -0.052 & 0.071 & -0.002 & 0.018 & 0.002 & 0.002 & 0.4 \\ 
		ACL2 & -0.052 & 0.071 & -0.002 & 0.018 & 0.002 & 0.001 & 0.4 \\ 
		ACL1 & -0.051 & 0.071 & 0.003 & 0.025 & 0.002 & 0.002 & 0.4 \\ 
		\hline
		variogNaive & -0.410 & 0.412 & 0.158 & 0.160 & 0.444 & 0.453 & 0.6 \\ 
		variogAdj & -0.055 & 0.072 & 0.024 & 0.041 & 0.007 & 0.011 & 0.6 \\ 
		geoNaive & -0.458 & 0.466 & 0.138 & 0.142 & 0.456 & 0.458 & 0.6 \\ 
		CL & -0.057 & 0.071 & 0.000 & 0.022 & 0.009 &  0.010 & 0.6 \\ 
		ACL2 & -0.057 & 0.071 & 0.000 & 0.022 & 0.009 & 0.010 & 0.6 \\ 
		ACL1 & -0.057 & 0.072 & 0.010 & 0.028 & 0.009 & 0.011 & 0.6 \\ 
		\hline
		variogNaive & -0.519 & 0.522 & 0.268 & 0.271 & 0.574 & 0.579 & 0.8 \\ 
		variogAdj & -0.067 & 0.092 & 0.037 & 0.049 & 0.030 & 0.041 & 0.8 \\ 
		geoNaive & -0.571 & 0.577 & 0.187 & 0.195 & 0.566 & 0.580 & 0.8 \\ 
		CL & -0.063 & 0.080 & -0.004 & 0.025 & 0.038 & 0.039 & 0.8 \\ 
		ACL2 & -0.063 & 0.080 & -0.004 & 0.025 & 0.038 & 0.039 & 0.8 \\ 
		ACL1 & -0.063 & 0.079 & 0.009 & 0.035 & 0.032 & 0.033 & 0.8 \\ 
		\hline
		variogNaive & -0.587 & 0.603 & 0.437 & 0.444 & 0.667 & 0.675 & 1.0 \\ 
		variogAdj & -0.066 & 0.097 & 0.051 & 0.060 & 0.023 & 0.030 & 1.0 \\ 
		geoNaive & -0.655 & 0.663 & 0.243 & 0.251 & 0.653 & 0.660 & 1.0 \\ 
		CL & -0.071 & 0.088 & -0.007 & 0.035 & 0.027 & 0.029 & 1.0 \\ 
		ACL2 & -0.071 & 0.088 & -0.008 & 0.031 & 0.025 & 0.028 & 1.0 \\ 
		ACL1 & -0.071 & 0.088 & 0.009 & 0.044 & 0.023 & 0.027 & 1.0 \\ 
		\hline
	\end{tabular}
\end{table}
		
\begin{table}[h]
	\renewcommand{\arraystretch}{1.15}
	\centering
\caption{\label{Tab:Gauss2}Bias and root-mean-square-error (RMSE) of the parameter estimates from the naive methods (variogNaive and geoNaive) and those accounting for positional error (variogAdj,  CL, ACL1 and ACL2). The true correlation function is Mat\'ern with $\kappa=1.5$.}
	\begin{tabular}{lccccccc}
		\hline
		& \multicolumn{2}{c}{$\sigma^2$} &  \multicolumn{2}{c}{$\phi$} &  \multicolumn{2}{c}{$\tau^2$} & $r=\delta/\phi$ \\ 
		\textbf{Method}  & Bias & RMSE & Bias & RMSE & Bias & RMSE \\ 
		\hline
		variogNaive & -0.049 & 0.095 & 0.009 & 0.023 & 0.051 & 0.073 & 0.2 \\ 
		variogAdj & -0.033 & 0.086 & 0.007 & 0.022 & 0.034 & 0.059 & 0.2 \\ 
		geoNaive & -0.053 & 0.079 & 0.008 & 0.012 & 0.048 & 0.051 & 0.2 \\ 
		CL & -0.038 & 0.084 & 0.000 & 0.011 & 0.035 & 0.059 & 0.2 \\ 
		ACL2 & -0.038 & 0.084 & 0.000 & 0.011 & 0.035 & 0.059 & 0.2 \\ 
		ACL1 & -0.038 & 0.085 & 0.002 & 0.012 & 0.035 & 0.060 & 0.2 \\ 
		\hline
		variogNaive & -0.124 & 0.154 & 0.018 & 0.030 & 0.124 & 0.141 & 0.4 \\ 
		variogAdj & -0.050 & 0.102 & 0.008 & 0.024 & 0.049 & 0.079 & 0.4 \\ 
		geoNaive & -0.147 & 0.160 & 0.020 & 0.024 & 0.141 & 0.145 & 0.4 \\ 
		CL & -0.052 & 0.099 & 0.001 & 0.011 & 0.049 & 0.078 & 0.4 \\ 
		ACL2 & -0.052 & 0.099&  0.001 & 0.011 & 0.049 & 0.078 & 0.4 \\ 
		ACL1 & -0.053 & 0.101 & 0.002 & 0.014 & 0.050 & 0.081 & 0.4 \\
		\hline 
		variogNaive & -0.212 & 0.230 & 0.031 & 0.044 & 0.217 & 0.228 & 0.6 \\ 
		variogAdj & -0.048 & 0.105 & 0.009 & 0.030 & 0.050 & 0.092 & 0.6 \\ 
		geoNaive & -0.243 & 0.252 & 0.035 & 0.039 & 0.240 & 0.244 & 0.6 \\ 
		CL & -0.051 & 0.104 & 0.000 & 0.013 & 0.049 & 0.086 & 0.6 \\ 
		ACL2 & -0.051 & 0.105 & 0.000 & 0.013 & 0.049 & 0.087 & 0.6 \\ 
		ACL1 & -0.052 & 0.107 & 0.002 & 0.019 & 0.050 & 0.090 & 0.6 \\
		\hline 
		variogNaive & -0.312 & 0.324 & 0.051 & 0.061 & 0.325 & 0.332 & 0.8 \\ 
		variogAdj & -0.059 & 0.118 & 0.013 & 0.032 & 0.066 & 0.114 & 0.8 \\ 
		geoNaive & -0.345 & 0.352 & 0.054 & 0.058 & 0.345 & 0.349 & 0.8 \\ 
		CL & -0.063 & 0.119 & 0.001 & 0.014 & 0.063 & 0.106 & 0.8 \\ 
		ACL2 & -0.063 & 0.119 & 0.001 & 0.014 & 0.063 & 0.106 & 0.8 \\ 
		ACL1 & -0.065 & 0.125 & 0.004 & 0.020 & 0.065 & 0.112 & 0.8 \\
		\hline 
		variogNaive & -0.400 & 0.408 & 0.079 & 0.090 & 0.420 & 0.426 & 1.0 \\ 
		variogAdj & -0.076 & 0.146 & 0.022 & 0.042 & 0.088 & 0.145 & 1.0 \\ 
		geoNaive & -0.433 & 0.438 & 0.074 & 0.078 & 0.433 & 0.435 & 1.0 \\ 
		CL & -0.082 & 0.144 & 0.001 & 0.021 & 0.083 & 0.134 & 1.0 \\ 
		ACL2 & -0.083 & 0.145 & 0.001 & 0.029 & 0.083 & 0.135 & 1.0 \\ 
		ACL1 & -0.086 & 0.152 & 0.006 & 0.022 & 0.087 & 0.148 & 1.0 \\ 
		\hline
	\end{tabular}
\end{table}

\section{Application}
\label{sec:application}

We analyse data on children's 
height-for-age Z-scores (HAZs) from a Demographic and Health Survey
\citep{Burgert2013-oq} conducted in Senegal in 2011. HAZs are a measure of the deviation from standard growth as defined by the WHO  Growth Standards \citep{deOnis2007} and are comparable across age and gender. A 
value of HAZ below -2 indicates stunted growth. \par
In this survey, the sampling unit are clusters of households within a 
predefined geographic area known as a census enumeration area (EA). An EA typically
corresponds to a single
city block or apartment building in urban areas and
to a village or group of
villages in rural areas The estimated centre of each cluster is recorded as a latitude/longitude coordinate, obtained from a GPS receiver or derived from public online maps or gazetteers \citep{Gething2015-fr}. To preserve the confidentiality of survey
respondents, uniform
geomasking was applied to the cluster centres. To take into account the different
population density, different values for the maximum displacement distance were applied to urban and rural locations, specifically $\delta_{urbarn}=2$ km
and $\delta_{rural}=5$ km. \par

The data consist of 384 clusters, of which 122 are urban, with 10 children per cluster on average. Our outcome of interest, $Y_{i}$, is the average HAZ for a cluster which we model as 
\begin{equation}
Y_i = \mu + S(x_i) + Z_i 
\label{eq:application}
\end{equation}
where $Z_i \sim N(0, \tau^2/n_i)$ and $n_{i}$ is the number of children at the $i$-th cluster. To account for positional error, we approximate uniform geomasking with its Gaussian counterpart as explained in Section \ref{subsec:uniform_geomasking}. \par
Table \ref{tab:results} reports the parameter estimates from the naive geostatistical model (first row) and the proposed modelling approach (second row). We were not able to obtain reliable estimates from the variogram-based correction approach due to the relatively high noise to signal ratio. As in simulation study, the naive geostatistical model approach yields a larger point estimates for $\tau^2$ and $\phi$ but smaller for $\sigma^2$. The point estimates for the practical range as estimated from for the naive model and
from
 the model that accounts for the geomasking are  $133.81$ km and $77.47$ km, respectively.  \par
 
 We also note that accounting for positional error also leads to narrower confidence intervals for all the model parameters expect the mean $\mu$. 

\begin{table}[h]
	\renewcommand{\arraystretch}{1.15}
	\centering
\caption{\label{tab:results} Parameter estimates and corresponding 95$\%$ confidence intervals (CI) for the fitted linear geostatistical models to malnutrition data of Section \ref{sec:application}. ``geoNaive'' is the naive approach which ignores positional error, while ``CL'' is the proposed approach based on the composite likelihood. }
	\begin{tabular}{rrcrc}
		\hline
		 & \multicolumn{2}{c}{geoNaive} & \multicolumn{2}{c}{CL} \\
Parameter & Estimate & 95$\%$ CI & Estimate & 95$\%$ CI  \\ 
		\hline
$\mu$ & -1.303 & (-1.470, -1.137) & -1.159 &  (-1.562, -0.736) \\
$\sigma^2$ &  0.117 & (0.045, 0.289) &  0.197 & (0.146, 0.257) \\ 
$\phi$ & 44.669 & (9.184, 80.138) & 25.860 & (17.782, 37.614)\\
$\tau^2$ & 0.536 & (0.081, 0.994) & 0.464 &  (0.409, 0.521) \\ 
		\hline
	\end{tabular}
\end{table}

\section{Discussion}

We have developed a computationally efficient approximation for maximum likelihood estimation of the linear geostatistical model under geomasking using the composite likelihood (CL) method. We have compared the performance of this approach with standard geostatistical approaches that ignore positional error and with a corrected variogram-based parametric fit. The CL method outperformed both, leading to substantially smaller root-mean-square-errors for the parameter estimates. The CL method
also provides a computationally more efficient alternative to \citet{Fanshawe2011-be} by reducing the computational burden from $O\left(B\times n^3 \right)$, where $B$ is the number of Monte Carlo simulations, to $O\left(n^2 \right)$. Our results indicate that ignoring positional error due to geomasking can lead to an overestimation of the nugget effect variance and the scale of the spatial correlation, while underestimating the variance of the residual spatial random effects. \par

The effects of geomasking on parameter estimation are stronger for larger values in the ratio $r = \delta/\phi$, where $\delta$ is the standard deviation of the positional error process in the case of Gaussian geomasking or the maximum displacement distance under uniform geomasking, $\phi$ is the scale of the spatial correlation. For this reason, geomasking procedures should always
 use the smallest acceptable value for $r$. High values of $r$ weaken the structure of the spatial dependence in the data, thus leading to less accurate predictive inferences. \par  

In the application of Section \ref{sec:application}, we approximated uniform geomasking with its Gaussian counterpart. As shown in Section \ref{subsec:uniform_geomasking}, the discrepancies between the true cumulative density function (CDF) of $[U_{ij}^* | u_{ij}]$ under uniform geomasking and the CDF of the Rice distribution that we used as an approximation are small in this case. Future research will investigate the 
robustness of this approach more generally under different scenarios. 

\bibliographystyle{apalike}
\nocite{*}
\bibliography{paper}

\appendix

\section{Appendix}

\subsection{Rice distribution}

The random variable $U$ follows a $Rice(\nu,\sigma)$ if its density function is
\[
f\left(u;\nu,\sigma\right)=\frac{u}{\sigma^{2}}\exp\left(-\frac{u^{2}+\nu^{2}}{2\sigma^{2}}\right)I_{0}\left(\frac{u\nu}{\sigma^{2}}\right),
\]
with $I_{k}\left(\cdot\right)$ is the modified Bessel function of
the first kind with order $k$. \par

The mean of $U$ is 
$$
E[U] = \sigma \sqrt{\frac{\pi}{2}} L(\nu^2/2\sigma^2)
$$
where
$$
L(x) = e^{x/2}\left[(1-x)I_{0}(x/2)-xI_{1}(x/2)\right];
$$
the variance is
$$
\text{Var}[U] = 2\sigma^2+\nu^2-\frac{\pi\sigma^2}{2}L^2(-\nu^2/2\sigma^2).
$$
A Rice variable is  also obtained as
$$
U = \sqrt{X_{1}^2+X_{2}^2},
$$
where $X_{1}$ and $X_{2}$ are independent Gaussian variables both with variance $\sigma^2$, and mean $\nu \cos \theta$ and $\nu \sin \theta$, respectively.
\end{document}